# Scattering of Optical Field by a Rough Multi-Scale Mirror Surface

P. A. Golovinsky, D. K. Proskurin

Voronezh State Technical University, Voronezh, Russia
E-mail: golovinski@bk.ru

Scattering of optical waves by a multi-scale rough mirror surface as a phase screen is considered. To solve the problem we used the diffusion phase approximation and numerical model of a phase jump. The scattering intensity was averaged over the ensemble of realizations. The results of numerical calculations of the angular dependence of the intensity are presented for a backscattering demonstrating the effect of the roughness distribution over the scales and its secondary impact on the indicatrix of scattering for multiplicative surface modulation.

The presence of two characteristic regions in the angular dependence of the intensity of light reflected by complex rough surface is reproduced in the two-scale model [1], but it has different scales clearly separated and scattering within each scale of inhomogeneity is described separately by the Kirchhoff tangent plane method or in the framework of the Rayleigh small perturbation approximation [2]. The two-scale model does not answer the question about wave scattering by rough surfaces with more complex roughness statistics. We consider the phase-modulating surface, which allows us to simplify the description of scattering in the presence of multi-scale inhomogeneities. We will consider a surface consisting of flat ideally mirror horizontal sections of different sizes and heights. When reflected from such a surface, the phase of the electric field in the electromagnetic wave changes to $\pi$, since there is no field inside the ideal conductor. Due to the different heights of the reflecting elements above the underlying surface, the wave phase is modulated, on which, at large distances, the changes caused by the different wave propagation distance to the observation point are superimposed. This model coincides with the local phase jump method [3,4]. In fact, we apply the Rytov approximation [5], in which the effect of modulation of the reflected wave is reduced to a random phase screen action. We take this into account in the scalar approximation for the electric component $E$ (s-polarization) of the laser field using the Kirchhoff-Helmholtz integral formula in the form [5]

$$E(\mathbf{R}_1) = -\frac{1}{4\pi} \int \frac{\partial E_S(\mathbf{r}, z)}{\partial z} \left( \frac{e^{ik|\mathbf{R}-\mathbf{R}_1|}}{|\mathbf{R}-\mathbf{R}_1|} \right) d\mathbf{r} \qquad (1)$$

Here $E_S(\mathbf{r})$ is the field taken directly near the reflecting surface, $\mathbf{R} - \mathbf{R}_1$ is radius vector directed from the current point of the reflecting surface $\mathbf{R}$ to the field reference point $\mathbf{R}_1$. For an incident wave with a unit amplitude $\exp(ik(\mathbf{n}_i^{\perp}\mathbf{r} + zn_i^z))$, the field of the reflected wave directly near the surface is equal

$$E'(\mathbf{r}) = -\exp(ik\mathbf{n}_i^{\perp}\mathbf{r})\exp[-ik\zeta(\mathbf{r})n_i^z] \qquad (2)$$

The expression for calculating the scattered wave in the far zone in the form of an integral over a stepped surface has the form

$$E(\mathbf{R}_1) = -\frac{ik_z^i}{2\pi}\int \exp(ik\mathbf{n}_i^{\perp}\mathbf{r} - ik\zeta(\mathbf{r})n_i^z)\frac{\exp\{i[kR_1 - k\mathbf{q}_s\mathbf{r} - k_z\zeta(\mathbf{r})]\}}{R_1}d\mathbf{r}, \qquad (3)$$

where $\mathbf{k} = \{\mathbf{q}_s, k_z\} = k(\mathbf{R}_1 / R_1) = k\mathbf{n}^s$, and $\mathbf{q}_s = k\mathbf{r}_1 / R_1$. When the reflecting surface is shifted by the magnitude $\zeta(\mathbf{r})$, the phase of the reflected wave changes by $\Delta\psi = -k\zeta(\mathbf{r})n_i^z$. We introduce the transverse component of the scattering vector $\mathbf{Q} = k(\mathbf{n}_i^{\perp} - \mathbf{n}_s^{\perp}) = \mathbf{q}_i - \mathbf{q}_s$ and the longitudinal component $K = k_i^z + k_z$. Then the average intensity of the field reflected from one-dimensional random roughness is

$$\bar{I} = \frac{k_z^{i2}S}{\pi^2 R_1^2}F_{\varsigma}(Q_x, K)\delta(Q_y), \qquad (4)$$

where

$$F_{\varsigma}(Q_x, K) = \frac{1}{2\pi}\int\left\langle\exp\{iK[\zeta(x) - \zeta(0)]\}\right\rangle\exp(iQ_x x))dx \qquad (5)$$

The condition $\delta(Q_y)$ in equation (4) means plane mirror reflection along the $y$ axis. Equation (5) gives a correlation function of harmonic oscillations with random phases, for which the corresponding theory is developed in statistical radiophysics [6]. In fact, the average over the ensemble $\left\langle\exp\{iK[\zeta(x) - \zeta(0)]\}\right\rangle$ can be calculated for specific distribution of magnitude $\zeta(x) - \zeta(0)$, for example, assuming that this is a normal process of diffusion type. Additional is the assumption that the deviation of roughness in the form $\Omega(x) = d\zeta(x)/dx$ is a stationary random process with $\langle\Omega(x)\rangle = 0$. The mean square phase shift is completely determined by the spectrum of random deviation of the wave roughness number $\Omega(x)$. We introduce the average square of fluctuations $\overline{\Omega^2}$. If $\overline{\Omega^2}l^2K^2 >> 1$, where $l$ is the correlation length, then the "height" $\overline{\Omega^2}l/K^2$ of the spectrum is much larger than its width $l$, the spectrum $g_{\Omega}(\kappa) = \int_{-\infty}^{\infty}\langle\Omega(x)\Omega(0)\rangle e^{i\kappa x}\,d\kappa / 2\pi$ is narrow, and the roughness varies greatly in magnitude on large spatial scales. For such a narrow spectrum, we have

$$F_\varsigma(Q,K) = \frac{1}{2\pi\sqrt{2\pi K^2 \overline{\Omega^2}}}\exp\left(-\frac{Q^2}{2K^2\overline{\Omega^2}}\right), \tag{6}$$

i.e. the Gaussian form of the spectral factor is obtained. In the second limiting case, when $\overline{\Omega^2}l^2K^2 \ll 1$, the surface roughness changes are characterized by small scales, and $g_\Omega(\kappa)$ range is wide. Then we are dealing with small-scale roughness,

$$F_\varsigma(Q,K) = \frac{g_\Omega(0)}{2}\frac{1}{\pi^2 g_\Omega(0)^2 K^2 + Q^2/K^2}\,, \tag{7}$$

and the scattering factor takes on the Lorentz form. Note that for both the wide and the narrow spectrum, the maximum scattering intensity corresponds to specular reflection ($Q = 0$). The conclusion about the dependence of scattering on the angle of incidence for different widths the deviation spectrum of roughness fluctuations is general in nature and is confirmed by direct numerical simulation [7].

In a more complex case, the surface is described by a random process, in which the deviation of the surface wavenumber $\Omega(x)$ is additionally modulated in a stepwise random manner from a large value to a small one and back on large scales, i.e. is an applicative surface. The scattering on such a surface critically depends on the statistics of roughness [8-10]. For complex statistics, the calculation of scattering can be performed by direct numerical simulation, integrating Eq. (5) with a partitioning to the rectangular shape surfaces with constant height, so that the structural factor takes the form

$$\int\exp[iQx + iK\zeta(x)]dx = \sum_m e^{i\psi_m}\frac{\sin(Ql_m/2)}{Q/2}\,, \tag{8}$$

where $\psi_m = \varphi_m + Q\left(\sum_{k=1}^m l_j - l_m/2\right)$, $\varphi_m = K\zeta_m$. The resulting wave of reflection will be represented by the sum of the waves with different amplitudes and phases, i.e. as a signal with amplitude-phase modulation. The phase of a separate component of the superposition of waves is made up of the currently modulation due to the height $\zeta_m$ of the step and the phase shift at an oblique incidence on the length $\sum_{k=1}^m l_j - l_m/2$ from the edge of the surface to the middle of the current strip with number $m$. The amplitude of a single component is determined by an oscillating factor, which sets the sharp directivity of the scattered wave at the angle of reflection for large values $l_m$, when $kl_m \gg 1$. In the opposite limit, when $kl_m \ll 1$, the scattering of a wave by a separate step loses its directionality, since $\sin(Ql_m/2)/(Q/2) \approx l_m$. Thus, the first factor in the sum expression of Eq. (8) is responsible for the coherence of the addition waves, and the second for the amplitude and direction of the individual components. When $Q \neq 0$ a progressive accumulation of phase occurs, making the whole process non-stationary, which reinforces the phase diffusion model described above.

Let us specify the statistical model of the applicative surface. A suitable type of random process is known in radio physics as a random fading signal [11]. Mathematically, the fading effect appears multiplicatively in the form

$$\zeta(x) = s_1(x)s(x),\qquad(9)$$

where $s_1(x)$ means a random variable slowly varying with distance, and $s(x)$ defines smaller random changes in the local relief.

The main random process determines the surface roughness in the absence of a smooth modulation. It can be set, for example, using the Rayleigh one-sided distribution [12] for step lengths

$$P(l) = \frac{l}{a_l^2}\exp\left(-\frac{l^2}{2a_l^2}\right),\ \langle l \rangle = (\pi/2)^{1/2}a_l,\ \langle x^2 \rangle = 2a_l^2 \qquad(10)$$

and independent distribution of their heights $\zeta$. The steps abut each other and $a_l >> a_\zeta$. This allows one to not take into account the rescattering without explicit limitation of the height of the relief. For the modulation coefficient of the relief $\beta$ ($a_\beta \sim 1$) and the intervals of the modulating $L$ slow process ($a_L >> a_l$) also uses the Rayleigh distribution. The choice of this distribution is dictated by the need to describe strictly positive values that is impossible using the Gaussian distribution, which is not limited on the number axis. The successful application the Rayleigh distribution to the description of fading in radio engineering [13] and to the parameterization of the distribution of various natural objects in size is well known.

For the numerical calculation of the intensity, we utilize its consistent estimate [14]

$$\bar{I} = \frac{k_z^2}{R_1^2}\frac{1}{n}\sum_{j=1}^{n}\frac{1}{X}\left|\int_{(j-1)X}^{jX}\exp[iQx + iK\zeta(x)]dx\right|^2,\qquad(11)$$

where $X >> l, L$. The Eq. (11) ensures that the dispersion tends to zero at $X \to \infty$, $n \to \infty$. The finiteness of the number of intervals leads to incomplete statistical averaging over possible implementations, and the finiteness of the interval length, i.e. the relative smallness of the wave scattering region as compared with the scale of roughness correlations ensures that the partial coherence of the scattered waves is preserved.

Fig. 1 shows an example of a random stepped surface profile with steps of different length and height.

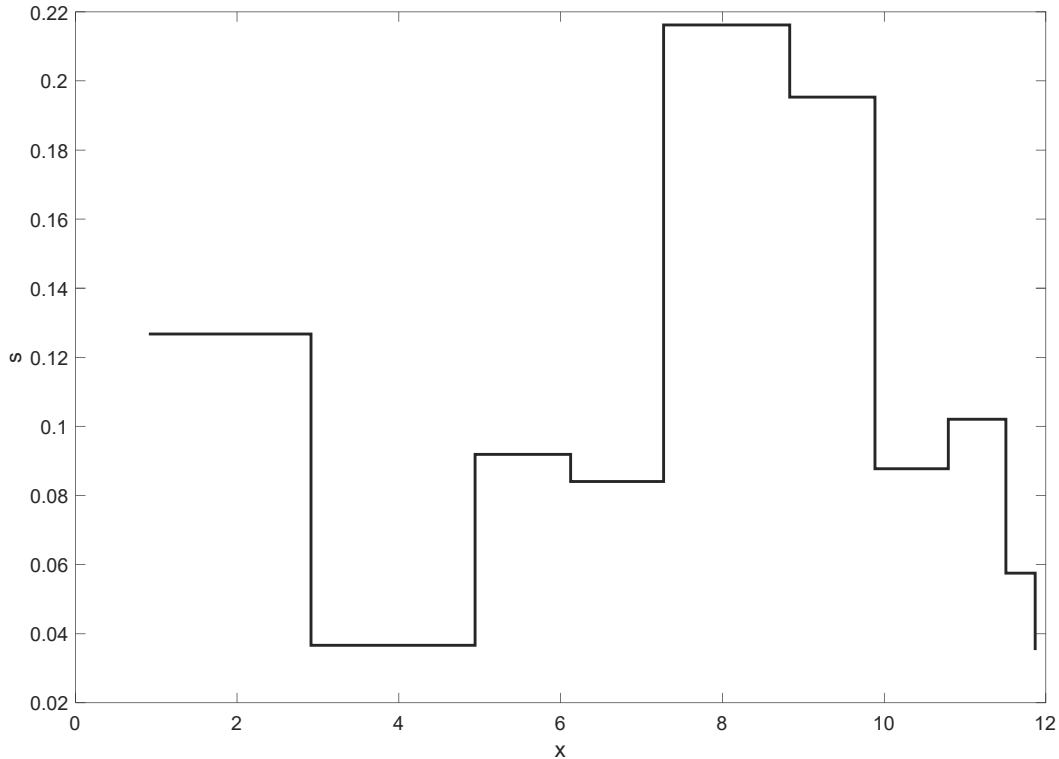

**Fig. 1.** The profile of a random stepped surface.

For such a surface, the scattering indicatrix was calculated with averaging the dependence over the ensemble of realizations. Similar calculations for $\lambda = a_l/2$ clearly reveal the tendency of broadening of the angular spectrum due to the growth of the diffraction divergence of the waves reflected from separate parts of the surface. This feature is even more visible when the parameter $\lambda = 2.2a_l$. The calculation results are presented in fig. 2.

When $\lambda = 2.2a_l$ a feature of dependence is manifested, it consists in the presence of a sharp maximum at small viewing angles and a hollow decreasing intensity at large angles. This duality is explained by the fact that in this distribution there are quite clearly represented areas with sizes both larger and smaller than the wavelength. For larger sizes, diffraction amplification in the $\theta \approx 0$ direction is manifested, while the areas of small length compared to the wavelength give diffuse scattering. This confirms the general qualitative conclusion made above on the basis of asymptotic analytical estimates that large-scale inhomogeneities give rise to an exponentially decaying spectrum, and small-scale inhomogeneities give a smoother Lorentz decrease in intensity. In the phase jump model, the kink in the dependence arises naturally due to significantly different diffraction conditions for large and small inhomogeneities.

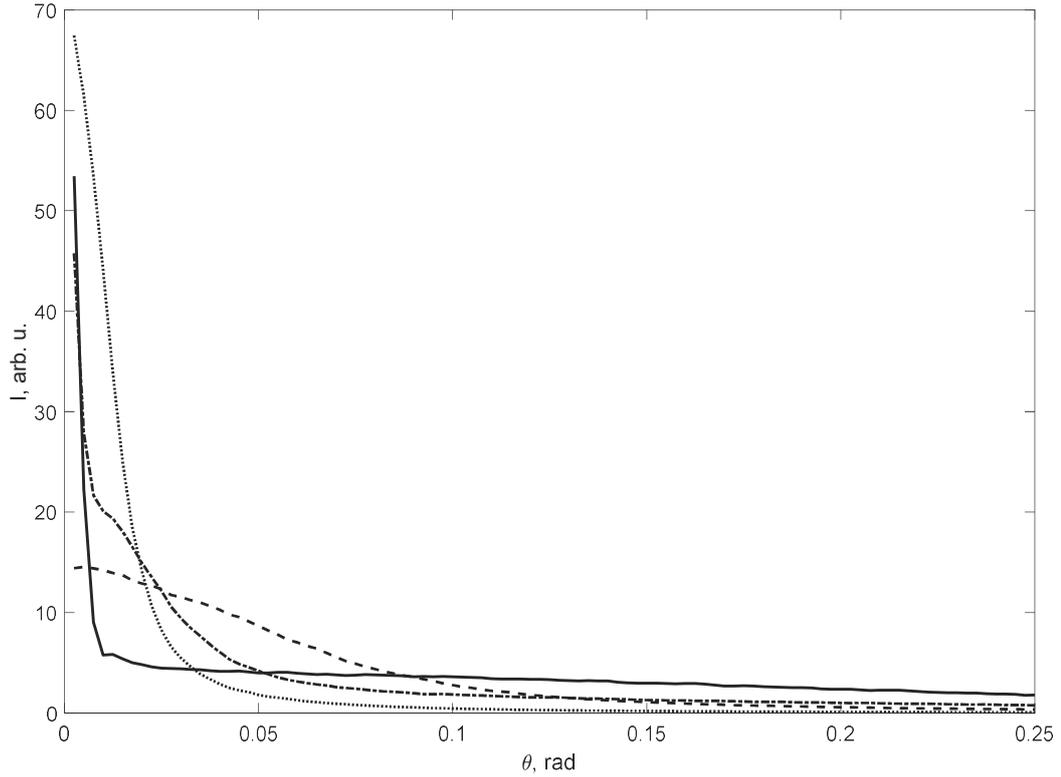

**Fig. 2.** Averaged over the realizations of a random surface, the dependence of the backscattering intensity on the angle: $\lambda = a_l/10$ - points, $\lambda = a_l/2$ - dashed line; $\lambda = 2.2a_l$ - solid line, dash-dotted line - the result of additional random large-scale modulation of roughness ( $a_L = 10a_l$ ).

Repeated, but already more large-scale random modulation of the surface, leads to the appearance of a fraction of areas of increased coherence, which is manifested in the redistribution of backscattered radiation to smaller angles. It is significant that randomly alternating degrees of roughness provide the possibility of additional control of the scattering process. The phase screen model makes it possible to determine variations in the efficiency of radiation scattering by a random one-dimensional stepped mirror surface with different statistical parameters of roughness, changing the final backscatter pattern.